\documentclass{article}
\usepackage{amssymb,amsmath,amscd}
\usepackage{graphicx,calc,epsfig,pstricks,bbm}
\usepackage{tikz}
\usepackage{authblk}
\usepackage[T1]{fontenc}
\usepackage[utf8]{inputenc}
\newcommand {\be}{\begin{equation}}
\newcommand {\ee}{\end{equation}}
\newcommand{\ba}{\begin{array}{c}}
\newcommand{\ea}{\end{array}}

\newcommand{\vgraph}{\mathfrak{n}}
\newcommand{\cube}{\ba
 \begin{tikzpicture}
\pgfmathsetmacro{\cubex}{0.15}
\pgfmathsetmacro{\cubey}{0.15}
\pgfmathsetmacro{\cubez}{0.15}
\draw (0,0,0) -- ++(-\cubex,0,0) -- ++(0,-\cubey,0) -- ++(\cubex,0,0) -- cycle;
\draw (0,0,0) -- ++(0,0,-\cubez) -- ++(0,-\cubey,0) -- ++(0,0,\cubez) -- cycle;
\draw (0,0,0) -- ++(-\cubex,0,0) -- ++(0,0,-\cubez) -- ++(\cubex,0,0) -- cycle;
\end{tikzpicture}
\ea}

\title{Quantum Reduced Loop Gravity}
\author[*]{Emanuele Alesci}%
\author[**]{Francesco Cianfrani}
\affil[*]{Instytut Fizyki Teoretycznej, Uniwersytet Warszawski, ul. Pasteura 5, 02-093 Warszawa, Poland, EU\\
        E-mail: emanuele.alesci@fuw.edu.pl}
\affil[**]{Institute for Theoretical Physics, University of Wroc\l{}aw, Plac\ Maksa Borna
9, Pl--50-204 Wroc\l{}aw, Poland.\\
        E-mail: francesco.cianfrani@ift.uni.wroc.pl}

\begin{document}
  \maketitle	

\abstract{Quantum Reduced Loop Gravity provides a promising framework for a consistent characterization of the early Universe dynamics. Inspired by BKL conjecture, a flat Universe is described as a collection of Bianchi I homogeneous patches. The resulting quantum dynamics is described by the scalar constraint operator, whose matrix elements can be analytically computed. The effective semiclassical dynamics is discussed, and the differences with Loop Quantum Cosmology are emphasized.}

%\FullConference{Frontiers of Fundamental Physics 14 - FFP14,\\
%                15-18 July 2014\\
%                Aix Marseille University (AMU) Saint-Charles Campus, Marseille }

\section{Introduction}

Early cosmology is a privileged arena in which our quantum theories of the gravitational field can be developed and, eventually, tested. The standard description of the global behavior of our Universe is given in terms of homogeneous and isotropic solutions of Einstein equations in the presence of matter field, {\it i.e.} FRW models. This description is appropriate in the present epoch at scales bigger than 100 Mpc, while at smaller scales the presence of structures (galaxies and cluster of galaxies) cannot be ignored. Going backward in time, the density contrast of the Universe is smaller and smaller, thus homogeneity and isotropy holds at more and more scales. The homogeneity and isotropy of the cosmic microwave background radiation spectrum provides the most convincing test for the standard cosmological model at recombination (when the averaged energy of the thermal bath is of the order of hydrogen atom binding energy). In earlier stages, inflation is expected to occur and it provides a partial explanation for homogeneity in later phases. Moreover, the fluctuation of the inflaton field during inflation are responsible for the patter of fluctuations observed in CMB spectrum. Hence, the inflationary scenario can be tested. 

If one goes further back in time and retains the assumptions of homogeneity and isotropy, then the initial Big Bang singularity is reached and at that time General Relativity is not predictive anymore. A more realistic scenario is realized skipping both homogeneity and isotropy. In fact, the causal horizon shrinks faster than the cosmological horizon towards the singularity, such that soon or later the Universe is a collection of not-causally-connected sub-Universes. The mathematical treatment of this phase is given via the BKL conjecture \cite{Belinsky:1982pk}, in which each sub-Universe is a proper homogeneous space, Bianchi type IX. However, each Bianchi IX space exhibits chaotic behavior towards the singularity. 

Therefore, the primordial phase of the Universe cannot be described in terms of the standard paradigm for cosmology, but it is expected to be the realm of Quantum Cosmology, which in turn is expected to descend from a complete Quantum Gravity theory. 

Loop Quantum Gravity (LQG) \cite{revloop} stands among the most promising quantum theories of gravity, even though a complete dynamic analysis is still missing. The standard cosmological implementation is Loop Quantum Cosmology (LQC) \cite{Bojowald:2011zzb}, which realizes the quantization of FRW models (minisuperspace quantization) using loop techniques. The main implication of LQC is the replacement of the initial singularity with a bounce. Furthermore, quantum corrections to the Universe dynamics affect the propagation of inflaton fluctuations during inflation and could provide testable modifications to the CMB spectrum \cite{Calcagni:2012vw}. 

An alternative description of the primordial Universe is given by Quantum-reduced Loop Gravity (QRLG) \cite{Alesci:2012md,Alesci:2013xd,Alesci:2013xya}. Since the homogeneity assumption cannot really hold on a quantum level, the idea is to provide a quantum description for a flat Universe as a collection of Bianchi I patches, along the same lines as when the BKL conjecture holds. This means restricting the metric to the inhomogeneous extension of Bianchi I model and then neglecting the interaction between different patches in the dynamic analysis. The latter corresponds to neglecting spatial gradients of the metric with respect to time derivatives and in the original BKL formulation this can be done close to the singularity. Here, the same assumption is expected to hold both close to the singularity (but a direct proof has not been given yet) and in the nearly-homogeneous case, in which spatial gradients are obviously negligible. 

Such a restriction on the metric is realized in the kinematical Hilbert space of LQG and, together with a gauge-fixing of the internal SU(2) group, it sets up the kinematical sector of QRLG. The dynamics is described by neglecting spatial gradients, which implies that the Hamiltonian is proportional to the euclidean scalar constraint, whose expression can be quantized such that the associated matrix elements can be computed analytically. The semiclassical limit of the Hamiltonian defines the effective dynamics, which is studied to infer the phenomenological implications of QRLG \cite{Alesci:2014uha,Alesci:2014rra}. 

\section{Kinematical Hilbert space}
QRLG provides a quantum framework for the investigation of the dynamics of the inhomogeneous Bianchi I model, which is described by the following line element
\begin{equation}\label{iB1}
ds^2=N^2(t)dt^2-a_1^2(t,x)(dx^1)^2-a_2^2(t,x)(dx^2)^2-a_3^2(t,x)(dx^3)^2\,,
\end{equation}
in the limit in which the spatial gradients of the scale factors $a_i$ ($i=1,2,3$) are negligible with respect to time derivatives (BKL-like limit). 

Furthermore, by gauge-fixing internal rotations the triads are set to be exactly diagonal, such that inverse densitized triads $E^a_i$ and Ashtekar connections are diagonal too, {\it i.e.}
\begin{align}
&E_i^a=(\ell_0)^{-2}p_i\delta_i^a\,,\qquad |p_1|=\ell_0^{2}\,a_2a_3\quad |p_2|=\ell_0^{2}\,a_3a_1\quad |p_3|=\ell_0^{2}\,a_1a_2\\
&A^i_a=(\ell_0)^{-1}c_i\delta^i_a+..\,,\qquad c_i=\frac{\gamma \ell_0}{N}\dot{a}_i\,,\label{Aia}
\end{align}  
$\gamma$ being the Immirzi parameter, while $\dot{a}_i$ denotes the time derivatives of $a_i$ and $\ell_0$ is the coordinated length of the considered spatial region. Indeed, $E^a_i$ can be exactly taken as diagonal, while $A^i_a$ generically contain some off-diagonal terms (the dots in \eqref{Aia}), related with the spatial spin connections and proportional to the spatial gradients of the scale factors, which can be neglected in the BKL limit. 

The associated Hilbert space is made of elements based at the cuboidal graphs $\Gamma$ whose links $l_i$ are parallel to the fiducial vectors $\vec{\delta}_i=\delta_i^a\partial_a$. At each link we deal with the following basis elements  
\be
{}^l\!D^{j_{l}}_{m_{l} m_{l}}(h_{l})=\langle j_l,m_l|\, R^{-1}_{l}D^{j_{l}}(h_{l}) R_{l}\,|j_l, m_l\rangle\qquad m_l=\pm j_l\,,
\ee
$D^{j_{l}}(h_{l})$ being Wigner matrices associated with the SU(2) group element based at $l$, while $R_{l}$ denotes the rotation mapping the tangent unit vector to $l$, $\vec{\delta}_l$, in $\vec{\delta}_3$. For $m_l=j_l$, $\vec{\delta}_l$ is choosen with the same orientation as $l$, while for $m_l=-j_l$ $\vec{\delta}_l$ has the opposite orientation with respect to $l$. 

A basis element based at a generic graph $\Gamma$ is constructed as follows
\be
\langle h|\Gamma, {\bf m_l, x_n \bf}\rangle= \prod_{n\in\Gamma}\langle{\bf j_{l}}, {\bf x}_n|{\bf m_{l}},  \vec{{\bf \delta}}_l \rangle 
\prod_{l} \;{}^l\!D^{j_{l}}_{m_{l} m_{l}}(h_{l}),\quad m_l=\pm j_l
\label{base finale}
\ee
where the products $\prod_{n\in\Gamma}$ and $\prod_{l}$ extend over all the nodes $n\in\Gamma$ and over all the links $l$ emanating from $n$, respectively. The intertwiners $\langle{\bf j_{l}}, {\bf x}_n|{\bf m_{l}},  \vec{{\bf u}}_l \rangle$ are merely coefficients which can be obtained by projecting the standard SU(2) intertwiner basis elements $|{\bf j_{l}}, {\bf x}_n\rangle$ in Livine-Speziale \cite{Livine:2007vk} coherent states $|{\bf m_{l}},  \vec{{\bf \delta}}_l \rangle=\Pi_l | m_{l},  \vec{\delta}_l \rangle$ based at the links $l$. 

The action of holonomy operators on basis elements is given by using U(1) recoupling theory at each link, while fluxes operators $\hat{E}_i(S^i)$ read the magnetic indexes at dual links $l=l_i$:
\be
\hat{E}_i(S^i){}^l\!D^{j_{l}}_{m_{l} m_{l}}(h_{l})= 8\pi\gamma l_P^2\, m_{l}\,{}^l\!D^{j_{l}}_{m_{l} m_{l}}(h_{l}) \qquad l_i\cap S^i\neq \oslash \,,\label{redei}
\ee
$l_P$ being Planck length. 

The residual symmetry on a kinematical level is reduced diffeomorphisms invariance, {\it i.e.} the invariance under those transformations acting along one of the fiducial directions having constant parameters along the others. It is imposed by constructing the associated s-knots according with the procedure adopted in LQG to implement background independence. 

\section{Quantum and effective dynamics} 

The classical Hamiltonian describing the dynamics of the metric \eqref{iB1} within the BKL approximation scheme is the sum over all points of that of Bianchi I model, for which the Euclidean scalar constraint is proportional to the Lorentzian one. Hence, the full Hamiltonian is proportional to the sum over all points of the Euclidean scalar constraint, written in terms of holonomies and fluxes of QRLG. The resulting expression can be quantized by considering a graph-dependent cubulation of the spatial manifold and the associated operators reads
\be
\hat{H}[\mathcal{N}]:= \frac{2i}{24\pi \gamma^3l^2_P}\sum_{\cube}\mathcal{N}(\vgraph) \,  \, \epsilon^{ijk} \,
   \mathrm{Tr}\Big[{}^{R}\hat{h}_{\alpha_{ij}} {}^{R}\hat{h}^{-1}_{s_{k}} \big[{}^R\hat{h}_{s_{k}},{}^{R}\hat{V}\big]\Big]\,\qquad C(m)=\frac{2i}{24\pi \gamma l^2_P}\,, 
   \label{Hridotto}
\ee
$\mathcal{N}(\vgraph)$ being the lapse function at the node $\vgraph$. The sum in the expression above extends over all the nodes and over all the cubulations. The holonomies $h$ are in the fundamental representation and they are based at links, which belong to the graph at which the states are based. In particular, $s_k$ and $\alpha_{ij}$ denote a link and a square of the graph emanating from the considered node. The operator $\hat{V}$ is the volume operator of the region dual to the node and from  \eqref{redei} one can see how it just reads the square root of the magnetic indexes along the fiducial directions. Therefore, the matrix elements of the Hamiltonian operator among basis elements \eqref{base finale} can be analytically computed and the whole dynamic problem can be solved in QRLG.    

The effective dynamics has been analyzed by discussing the dynamics generated by the expectation value of $\hat{H}$ over semiclassical states. These semiclassical states have been constructed starting from coherent states at each link $l$ as follows
\begin{equation}
\psi^{{\bf\alpha}}_{\Gamma{\bf H'}}=\sum_{{\bf m_{l}}}\prod_{n\in\Gamma} \langle{\bf j_{l}}, {\bf x}_n|{\bf m_{l}},  \vec{{\bf u}}_l \rangle^*\;\prod_{l\in\Gamma} \psi^\alpha_{H'_{l}}(m_{l})\;\langle h|\Gamma, {\bf m_l, x_n \bf}\rangle\,,
\label{semiclassici ridotti inv}
\end{equation}
in which the functions $\psi^\alpha_{H'_{l}}(m_{l})$ are labeled by $H'_l=h_le^{\frac{\alpha}{8\pi\gamma l_P^2}E'_l\tau_l}$ and they are peaked around the classical values $h_l$ and $E_l$ for the holonomy and the dual flux $E_l$ along the links $l$ of the graph $\Gamma$.

The semiclassical analysis provides the following expectation value for the Hamiltonian operator (we neglected the correction due to the finite spread of the semiclassical wave packet)
\begin{align}
\langle\, \hat{H}\,\rangle_{N} \,\approx
\frac{2}{\gamma^2}\mathcal{N}\bigg(&N_1\, N_2\,\sqrt{p^1\;p^2}\,\frac{\sqrt{p^3+p^3_0}-\sqrt{p^3-p^3_0}}{p^3_0}\;  \sin{\frac{c_1}{N_1}} \sin{\frac{c_2}{N_2}}+\nonumber\\
&+N_2\, N_3\,\sqrt{p^2\;p^3}\,\frac{\sqrt{p^1+p^1_0}-\sqrt{p^1-p^1_0}}{p^1_0}\;  \sin{\frac{c_2}{N_2}} \sin{\frac{c_3}{N_3}}+\nonumber\\
&+N_3\, N_1\,\sqrt{p^3\;p^1}\,\frac{\sqrt{p^2+p^2_0}-\sqrt{p^2-p^2_0}}{p^2_0}\;  \sin{\frac{c_3}{N_3}} \sin{\frac{c_1}{N_1}}\bigg).
\end{align}
where 
\be\label{p0}
p^i_0=4\pi\gamma \frac{N}{N_i} l_P^2\,, 
\ee
$N_i$ being the total number of nodes along the fiducial direction $i$ within each homogeneous patch. It is worth noting how for $p^i\gg p^i_0$ and $c_i\ll N_i$ the leading order term coincides with the Hamiltonian for the inhomogeneous extension of the Bianchi I model, thus reconciling our quantum formulation with the classical limit. 

We see how generically there are two kind of corrections. Holonomy corrections are due the terms $N_i\sin{\frac{c_i}{N_i}}$. They retain the same form as the analogous ones in LQC as soon as one identifies the regulator $\mu_i$ with the inverse numbers of nodes along the direction $i$, {\i.e.} $\mu_i=1/N_i$. Since in LQC such corrections are responsible for the bounce, our effective semiclassical dynamics predicts a bouncing scenario too. 

Inverse volume corrections comes from the next-to-the-leading-order terms in the expansion of the functions $\frac{\sqrt{p^i+p^i_0}-\sqrt{p^i-p^i_0}}{p^i_0}$ for $p^i\gg p^i_0$. In particular, one gets
\be
\frac{\sqrt{p^i+p^i_0}-\sqrt{p^i-p^i_0}}{p^i_0}=\frac{1}{\sqrt{p^i}}\left(1+\frac{1}{8}\left(\frac{p^i_0}{p^i}\right)^2+O\left(\frac{p^i_0}{p^i}\right)^4\right)\,,
\ee
from which one sees how the corrections are enhanced with respect to those of LQC, because of the factor $N$ in the right-hand side of \eqref{p0}.

\section{Conclusions}

QRLG provides a framework for the quantum description of the early Universe. The quantum scalar constraint within the adopted approximation scheme can be analytically defined. The effective dynamics resembles that of LQC. In particular, holonomy corrections coincide once the regulators are identified with the numbers of nodes, as in lattice refinement \cite{Bojowald:2011iq}, while inverse volume corrections are enhanced. Therefore, the replacement of the initial sigularity with a bounce is still predicted, while stronger corrections are expected to occur when analyzing the behavior of perturbations.

Furthermore, since in QRLG one has a complete quantum description of the Universe, not restricted to minisuperspace, one can use loop techniques in the quantization of the matter sector. As a consequence, additional modifications with respect to LQC are to be found in the presence of matter fields. These will be the subject of forthcoming investigations.\\\\\\

\thanks{The work of FC was supported by funds provided by the National Science Center under the agreement DEC12
2011/02/A/ST2/00294.
The work of E.A. was supported by the grant of Polish Narodowe Centrum Nauki nr 2011/02/A/ST2/00300.}


\begin{thebibliography}{99}



\bibitem{Belinsky:1982pk} 
  V.~A.~Belinsky, I.~M.~Khalatnikov and E.~M.~Lifshitz,
  %``A General Solution of the Einstein Equations with a Time Singularity,''
  Adv.\ Phys.\  {\bf 31}, 639 (1982).	

\bibitem{revloop}
C. Rovelli, ``Quantum gravity'', (Cambridge University Press, Cambridge, 2004); 

A.~Ashtekar and J.~Lewandowski, Class.\ Quant.\ Grav.\  {\bf 21}, R53 (2004);

T. Thiemann, ``Modern Canonical Quantum General Relativity'', (Cambridge University Press, Cambridge, 2006).

\bibitem{Bojowald:2011zzb} 
  M.~Bojowald,
  %``Quantum cosmology : A Fundamental Description of the Universe,''
  Lect.\ Notes Phys.\  {\bf 835}, pp.1 (2011);

%\bibitem{Ashtekar:2011ni} 
  A.~Ashtekar and P.~Singh,
  %``Loop Quantum Cosmology: A Status Report,''
  Class.\ Quant.\ Grav.\  {\bf 28}, 213001 (2011)
  [arXiv:1108.0893 [gr-qc]].  

%\cite{Calcagni:2012vw}
\bibitem{Calcagni:2012vw} 
  G.~Calcagni,
  %``Observational effects from quantum cosmology,''
  Annalen Phys.\  {\bf 525}, no. 5, 323 (2013)
  [Erratum-ibid.\  {\bf 525}, no. 10-11, A165 (2013)]
  [arXiv:1209.0473 [gr-qc]].

%\cite{Alesci:2012md}
\bibitem{Alesci:2012md} 
  E.~Alesci and F.~Cianfrani,
  %``A new perspective on cosmology in Loop Quantum Gravity,''
  Europhys.\ Lett.\  {\bf 104}, 10001 (2013)
  [arXiv:1210.4504 [gr-qc]].
  %%CITATION = ARXIV:1210.4504;%%
	
	%\cite{Alesci:2013xd}
\bibitem{Alesci:2013xd} 
  E.~Alesci and F.~Cianfrani,
  %``Quantum-Reduced Loop Gravity: Cosmology,''
  Phys.\ Rev.\ D {\bf 87}, no. 8, 083521 (2013)
  [arXiv:1301.2245 [gr-qc]].

%\cite{Alesci:2013xya}
\bibitem{Alesci:2013xya} 
  E.~Alesci, F.~Cianfrani and C.~Rovelli,
  %``Quantum-Reduced Loop-Gravity: Relation with the Full Theory,''
  Phys.\ Rev.\ D {\bf 88}, 104001 (2013)
  [arXiv:1309.6304 [gr-qc]].

%\cite{Alesci:2014uha}
\bibitem{Alesci:2014uha} 
  E.~Alesci and F.~Cianfrani,
  %``Quantum Reduced Loop Gravity: Semiclassical limit,''
  Phys.\ Rev.\ D {\bf 90}, no. 2, 024006 (2014)
  [arXiv:1402.3155 [gr-qc]].
	
%\cite{Alesci:2014rra}
\bibitem{Alesci:2014rra} 
  E.~Alesci and F.~Cianfrani,
  %``Loop Quantum Cosmology from Loop Quantum Gravity,''
  arXiv:1410.4788 [gr-qc].
	
	%\cite{Livine:2007vk}
\bibitem{Livine:2007vk}
  E.~R.~Livine and S.~Speziale,
  %``A New spinfoam vertex for quantum gravity,''
  Phys.\ Rev.\ D {\bf 76},  084028 (2007)
  [arXiv:0705.0674 [gr-qc]].

%\cite{Bojowald:2011iq}
\bibitem{Bojowald:2011iq} 
  M.~Bojowald, G.~Calcagni and S.~Tsujikawa,
  %``Observational test of inflation in loop quantum cosmology,''
  JCAP {\bf 1111}, 046 (2011)
  [arXiv:1107.1540 [gr-qc]].

\end{thebibliography}
\end{document}